\documentclass[superscriptaddress,twocolumn,aps,pre,showpacs]{revtex4}
\usepackage{epsfig}
\usepackage{graphicx}
\usepackage{dcolumn}
\usepackage{bm}
\begin{document}

\title{Slow L\'evy flights}

\author{Denis Boyer}
\email{boyer@fisica.unam.mx}
\affiliation{Instituto de F\'\i sica, Universidad Nacional Aut\'onoma de 
M\'exico, D.F. 04510, M\'exico}
\affiliation{Centro de Ciencias de la Complejidad, Universidad Nacional 
Aut\'onoma de M\'exico, D.F. 04510, M\'exico}
\affiliation{Max-Planck-Institut f$\ddot{u}$r Physik komplexer Systeme, 
N$\ddot{o}$thnitzer Str. 38, D-01187 Dresden, Germany}

\author{Inti Pineda}
\affiliation{Instituto de F\'\i sica, Universidad Nacional Aut\'onoma de 
M\'exico, D.F. 04510, M\'exico}

\date{\today}

\begin{abstract}
Among Markovian processes, the hallmark of
L\'evy flights is superdiffusion, or faster-than-Brownian dynamics. Here
we show that L\'evy laws, as well as Gaussians, can also be
the limit distributions of processes with long range
memory that exhibit very slow diffusion, logarithmic in time.
These processes are path-dependent and anomalous motion emerges from 
frequent relocations to already visited sites. We show how the Central 
Limit Theorem is modified in this context, keeping
the usual distinction between analytic and non-analytic 
characteristic functions. A fluctuation-dissipation 
relation is also derived. Our results may have
important applications in the study of animal and human
displacements.
\end{abstract}

 \maketitle

\section{Introduction}

L\'evy flights (LFs) represent one of the most 
important extensions of the Central Limit Theorem (CLT), a cornerstone
of probability theory \cite{feller,alpha}.
LFs are sums of independent and identically 
distributed random variables that admit non-Gaussian limit laws due
to their very large fluctuations. They find physical 
applications in laser cooling \cite{bardou}, optics \cite{light}
or chaotic transport \cite{strangekin}. 
LFs are also paradigmatic of superdiffusive processes, {\it i.e.},
anomalous types of transport where the characteristic diffusive 
length scale $l(t)$ of an individual particle 
grows with time as $t^{\alpha}$ with $\alpha>1/2$, that is,
faster than in the classical Brownian 
motion (BM) \cite{bouchaud,rainer,klafterphysrep,weiss}.

In recent years, LFs (as well as the related L\'evy walks \cite{lw}) have 
become prominent for modeling diffusion in a variety of complex systems.
Power-law distributions of step lengths with diverging variance, a key
feature of L\'evy processes, are found to describe well the trajectories 
of immune cells in the brain \cite{tcells}, 
the displacements of animals \cite{gandhi,levy2,sims,gab} and 
hunter-gatherers \cite{brown,pnas2014} in their environments, 
or the travels of modern humans within and between
cities \cite{geisel,rhee,gonzalez,song}.  However, the assumption of 
independence between steps does limit the applicability of genuine 
L\'evy processes for modeling real systems, where 
non-Markovian effects and correlations can be strong. Deeper analysis of
empirical data actually reveals that the diffusion of humans and animals 
(even those exhibiting L\'evy patterns)
is in general {\it sub}diffusive at large times, {\it i.e.},
with $l(t)\ll t^{1/2}$ 
\cite{song,gautestad2005,gautestad2006,borger,solis}. 
Furthermore, $l(t)$ commonly grows more slowly than a power-law 
of time, namely, in a logarithmic way \cite{borger,song,solis}: 
this behavior is even in sharper contrast with 
the superdiffusion of simple LFs.

Logarithmic diffusion can be generated in several ways, for instance, 
by continuous time random walks models with superheavy-tailed distributions 
of waiting times \cite{kantz}, or by certain iterated 
maps \cite{drager,sokolov}. In the context of animal and human mobility, an 
important but little explored mechanism that may lead to very slow 
subdiffusion is spatial memory: many living organisms actually keep revisiting 
familiar places \cite{fagan,gautestad2005,gautestad2006,vanmoorter,borger,solis}. 
Here, we seek to understand, with the help of a solvable model, how this type 
of memory can act as a self-attracting force which drastically constrains
diffusion towards limited areas, giving rise to \lq\lq home ranges", and 
how this property can still be compatible with power-law distributed 
step lengths. 

The dynamics and limit distributions of constrained LFs 
are not well understood, except for
processes subjected to long waiting times or in external potentials, mainly 
\cite{klafterphysrep,compet}. Several limit theorems also exist 
for specific problems of sums of correlated random variables \cite{hilhorst}, 
and a few random walks with infinite memory of their previous displacements 
have exactly solvable first moments 
\cite{elephant,kumar,dasilva}. Yet, very little is known on 
LFs composed of non-independent steps, in particular processes with 
self-attraction.
Self-attracting random walks are path-dependent processes where 
a walker tends to return to previously visited sites 
\cite{davis,annals}. Numerical simulations 
and scaling arguments clearly show that self-attracting walks can exhibit
subdiffusion \cite{quasistatic,grassberger,siam}.
These mathematically challenging processes cannot be readily
analyzed with better known frameworks for subdiffusive phenomena, 
such as fractional Fokker-Planck equations \cite{klafterphysrep} or 
scaled Brownian motions \cite{sbm,slowsbm}. They are more related with 
diffusion in quenched disordered media \cite{bouchaud}, 
where some rigorous connections have been made with the
Sinai model \cite{matteo}.

In this study, we heuristically modify the CLT for processes that exhibit 
very slow diffusion, and show that such modification exactly describes 
a class of
self-attracting LF and self-attracting random walks. The characteristic 
functions having a similar structure than in the ordinary CLT, Gaussian and 
L\'evy distributions emerge asymptotically in space, although the dynamics 
is strongly subdiffusive. We also derive a fluctuation-dissipation 
relation in the Gaussian case.

\section{General formulation} 

Let $P(n,t)$ be the probability that the 
position $X_t$ of a particle at time $t$ is $n$ (where $n$ and $t$ are 
discrete), given that 
the particle is located at the origin $n=0$ at $t=0$.
We consider discrete, one dimensional walks, keeping in
mind that discreteness is not relevant in the asymptotic limit. 
The results can also be
extended higher dimensions straightforwardly.

We recall that for a standard random walk composed of $t$ i.i.d.
displacements $\ell_i$ with distribution $p(\ell)$, 
the characteristic function of $X_t$, defined as 
$\widetilde{P}(k,t)\equiv\sum_{n=-\infty}^{\infty}e^{-ikn}P(n,t)=
\langle e^{-ikX_t}\rangle$, takes the form \cite{weiss}:
\begin{equation}\label{exp}
\widetilde{P}(k,t)=\tilde{p}(k)^t=e^{\ln[\tilde{p}(k)]t},
\end{equation}
where $\tilde{p}(k)$ is the characteristic function of $\ell$. 
Since $\tilde{p}(0)=1$ by normalization, in the unbiased 
($\langle \ell\rangle=0$) and symmetric case, an expansion near $k=0$ gives:
\begin{equation}\label{pk}
\tilde{p}(k)=1-C|k|^{\mu}+... 
\end{equation}
Two basic situations emerge: 
the analytic case $\mu=2$, corresponding to $\langle \ell^2\rangle<\infty$
(and $C=\langle\ell^2\rangle/2$), and the non-analytic case $0<\mu<2$
when $\langle \ell^2 \rangle$ does not exist, due to a power-law decay
of $p(\ell)$:
\begin{equation}
p(\ell)\sim 1/|\ell|^{1+\mu} 
\end{equation}
at large $\ell$ \cite{weiss}. 
Combining (\ref{exp})-(\ref{pk}) yields the celebrated Gaussian-L\'evy CLT: 
\begin{equation}\label{tcl}
\widetilde{P}(k,t)\rightarrow e^{-C|k|^{\mu}t}.
\end{equation}
Eq. (\ref{tcl}) implies a scaling law 
$P(n,t)\rightarrow t^{-1/\mu}f(n/t^{1/\mu})$
where the scaling function $f(x)$ is a Gaussian
or a symmetric L\'evy law $L_{\mu,0}(x)$, for $\mu=2$ and $0<\mu<2$, 
respectively. The latter case is superdiffusive as the typical
diffusion length is $\propto t^{1/\mu}\gg t^{1/2}$.

Consider now a simple modification of Eq. (\ref{exp}):
suppose that for certain diffusion processes with memory
or sums of correlated random variables
(we do not need to specify a model at this point),
$\widetilde{P}$ is not an exponential function of $t$
but a {\it power-law}:
\begin{equation}\label{powerlaw}
\widetilde{P}(k,t)\simeq t^{-a(k)}=e^{-a(k)\ln t},
\end{equation}
at large $t$ and small $k$. The function
$a(k)$ satisfies $a(0)=0$, owing to the
normalization $\widetilde{P}(k=0,t)=1$.
Again, $a(k)$ can be generically analytic or non-analytic
near $k=0$. In the first case, since
$\widetilde{P}(k,t)^*=\widetilde{P}(-k,t)$ and 
$|\widetilde{P}(k,t)|\le 1$, the Taylor expansion of the exponent
must be of the form $a(k)\simeq ia_1 k+a_2 k^2+...$, 
with $a_1$ and $a_2$ two real constants and $a_2>0$. 
For simplicity, we first consider $a_1=0$, 
or motion without bias. 

In the non-analytic case, the same arguments lead to $a(k)\simeq
a_{\mu}|k|^{\mu}$ with $0<\mu<2$ {\it a priori}, and $a_{\mu}>0$. Inserting 
into (\ref{powerlaw}), we see that the main difference with (\ref{tcl}) 
is that the variable $t$ is substituted by $\ln t$. Hence:
\begin{equation}\label{loglevy}
P(n,t)\rightarrow \frac{1}{(\ln t)^{1/\mu}}\ 
f_{\mu}\left(\frac{n}{(\ln t)^{1/\mu}}\right),
\end{equation}
where the limit laws $f_{\mu}(x)$ are the same as in the ordinary CLT.
If $\mu=2$, diffusion is Gaussian but very slow: 
$\langle X_t^2\rangle=2a_2\ln t$, in sharp contrast with BM, 
where $\langle X_t^2\rangle=2D t$. 
[In this case, Eq. (\ref{loglevy})  should not be confused 
with the log-normal distribution, where the logarithm applies 
to the space variable, not the temporal one.] A basic Markovian example
is, by construction, a scaled Brownian motion, which is a BM where 
the time $T$ is rescaled as $t=e^T$. Such process is also equivalent 
to a BM with a time-dependent diffusion coefficient,
$D(t)$, decaying as $1/t$ at large $t$ \cite{slowsbm}.

In the non-analytic case, the situation looks paradoxical at first sight.
The ensemble average $\langle X_t^2\rangle=\infty$ like in ordinary 
L\'evy processes due to the broad tails of $L_{\mu,0}(x)$
(or due to the fact that $\partial^2P(k,t)/\partial k^2$ 
does not exist at $k=0$, from Eq.(\ref{powerlaw})). Yet, 
Eq. (\ref{loglevy}) also defines a typical diffusion length 
$l(t)\propto (\ln t)^{1/\mu}$, which grows extremely slowly.
Therefore, based on this scaling length $l(t)$, motion is strongly 
subdiffusive and 
all the finite moments, $\langle |X_t|^{\nu}\rangle$ with $\nu<\mu$, 
also evolve very slowly, as $(\ln t)^{\nu/\mu}$. Still, the
process keeps superdiffusive features through the divergence of the 
second moment.
This situation is reminiscent of scaling violation, which also arises
in continuous time random walks \cite{schmiedeberg} or 
L\'evy walks \cite{lw,fleurov}. 

\begin{figure}
  \centerline{ 
  \includegraphics*[width=0.37\textwidth,angle=-90]{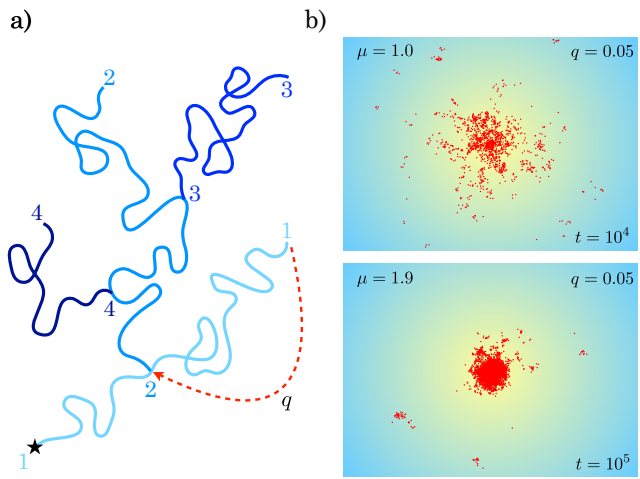}
  }  
  \caption{(Color online) {\bf a)} Schematic view of
  a process relocating at a constant 
  rate ($q$) to sites occupied at previous times, these times being chosen 
  stochastically. The numbers label the beginning and end of each 
  excursion. Each end is followed by the beginning of the next 
  excursion (arrow). {\bf b)} Two simulated
  trajectories corresponding to L\'evy excursions with
  $p(\ell)\sim 1/|\ell|^{1+\mu}$, relocation rate $q=0.05$
  and relocation kernel given by Eq. (\ref{pref}) 
  [panels at the same scale].
}
\label{fig1}
\end{figure}

\section{Random walks with relocations} 

We now consider a
concrete class of non-Markovian walks for which the above 
ideas apply. The processes of interest are  
self-attracting, namely, they tend to revisit locations visited 
in the past. Particular examples were studied numerically in 
\cite{gautestad2005,gautestad2006} as animal movement models, 
or theoretically in \cite{solis,romo}. We present here a unified view of
this class of processes.

Let $q$ be a parameter ($0<q<1$). At any time $t$, the walker chooses 
its next position according to the following rules: 

{\it (i)} with probability $1-q$, it performs a random displacement 
$\ell$ drawn from a given distribution $p(\ell)$ like in
standard random walks or L\'evy flights; 

{\it (ii)} with the complementary
probability $q$, it jumps (or 'reset') directly to the site occupied 
at some previous time $t'\le t$. The time $t'$ is chosen according to a
given probability $\pi_t(t')$, or memory function, 
with $\sum_{t'=0}^{t}\pi_t(t')=1$ by normalization.

The rules are depicted in Fig. \ref{fig1}a, with two simulated examples
in Fig. \ref{fig1}b.
Note that in {\it (ii)}, the next target site is chosen independently of its
distance to the location $X_t$ of the walker. If $\pi_{t}(t')=\delta_{t',0}$,
the site chosen for revisit is unique (the origin), a case which corresponds
to the well-studied random walk with resetting to the origin
\cite{evansmaj,montero,resetlevy,nowak}. 
For more general kernels, the walk is strongly path-dependent but still 
described by a master equation:
\begin{equation}\label{master}
P(n,t+1)=(1-q)\sum_{\ell=-\infty}^{\infty}p(\ell)P(n-\ell,t)
+q\sum_{t'=0}^t \pi_t(t')P(n,t').
\end{equation}
Standard random walks or L\'evy flights are recovered for $q=0$. If $q\ne0$,
the last term indicates that site $n$ can be chosen to be 
occupied at time $t+1$, provided it was visited at the earlier time $t'$. 

\begin{figure}
  \centerline{ 
\hspace{-0.4cm}  \includegraphics*[width=0.39\textwidth,angle=-90]{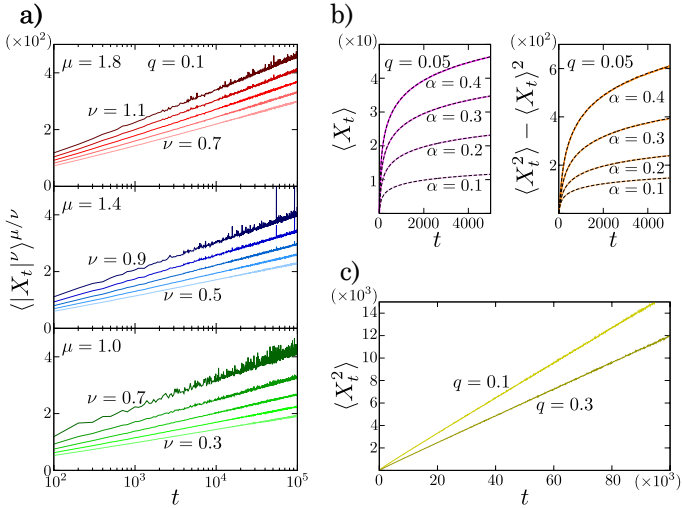}
  }  
  \caption{(Color online) Preferential visit model in $1d$. {\bf a)}
  $\langle |X_t|^{\nu}\rangle^{\mu/\nu}$, obtained from simulations
  with different $\mu$ and $\nu$ (averages over $5\times 10^5$ runs), 
  is proportional to $\ln t$ as expected.
  {\bf b)} Mean and variance of $X_t$ for a n.n. walk
  with bias $\alpha$ in
  rule {\it(i)}. Colored solid lines are simulations and dark dashed lines, 
  theory. {\bf c)} Normal diffusion for spatially uniform relocations.
}
\label{fig2}
\end{figure}

We first consider a uniform memory function, that is, independent of $t'$:
\begin{equation}\label{pref}
\pi_t(t')=\frac{1}{t+1}.
\end{equation} 
We call this case the {\it preferential visit model} (PVM): 
with such kernel, rule {\it (ii)} is simply equivalent to choosing a given 
site $n$ (among all visited sites) with probability proportional to
the number of visits received by $n$ since $t=0$.
Therefore the walker is prone to revisit familiar sites, at the expanse
of rarely visited ones. The moments $\langle X_t^{2p}\rangle$ where
calculated in \cite{solis} for the PVM with nearest neighbor (n.n.) 
steps ($\ell_i=\pm1$) in rule {\it (i)}. To solve Eq. (\ref{master}) 
more generally,
we define the Laplace transform of $\widetilde{P}(k,t)$:
\begin{equation}
\widehat{P}(k,\lambda)=\sum_{t=0}^{\infty}\lambda^t
\sum_{n=-\infty}^{\infty} e^{-ikn}P(n,t).
\end{equation}
By taking the double transform of Eq. (\ref{master}) with the
kernel (\ref{pref}) and writing
$\lambda^t/(t+1)=\lambda^{-1}\int_0^{\lambda}u^tdu$, we obtain:
\begin{equation}\label{FL1}
\widehat{P}(k,\lambda)-1=(1-q)\tilde{p}(k)\lambda\widehat{P}(k,\lambda)
+q\int_0^{\lambda}du\frac{\widehat{P}(k,u)}{1-u}.
\end{equation}
Taking the derivative of Eq. (\ref{FL1}), one obtains a first-order 
ODE in the variable $\lambda$. As
$P(n,t=0)=\delta_{0,n}$, the condition
$\widehat{P}(k,0)=1$ must be enforced, leading to the exact solution:
\begin{equation}\label{solution}
\widehat{P}(k,\lambda)=(1-\lambda)^{-[1-a(k)]}
\left[1-(1-q)\tilde{p}(k)\lambda\right]^{-a(k)}
\end{equation}
with
\begin{equation}\label{a}
a(k)=(1-q)\frac{1-\tilde{p}(k)}{1-(1-q)\tilde{p}(k)}.
\end{equation}
We can infer the large $t$  behavior of $\widetilde{P}(k,t)$ 
by studying the divergence of $\widehat{P}(k,\lambda)$ near
$\lambda=1$, with $k$ fixed but small. Noting that
$a(k)\ll 1$, Eq. (\ref{solution}) yields
$\widehat{P}(k,\lambda)\simeq (1-\lambda)^{-[1-a(k)]}$.
This expression is simply inverted as:
\begin{equation}\label{pzt}
\widetilde{P}(k,t)\simeq t^{-a(k)},
\end{equation}
as announced in (\ref{powerlaw}). In the absence of bias, one can use 
Eq. (\ref{pk}), which, combined with (\ref{a}), gives the exponent: 
\begin{equation}\label{aasym}
a(k)\simeq \frac{1-q}{q}C|k|^{\mu}, 
\end{equation}
implying the limit law (\ref{loglevy}). We conclude that this
random walk always diffuses logarithmically, unlike other re-inforced walks
that exhibit transitions to localized states \cite{davis,grassberger}.
Numerical simulations confirm the 
very slow dynamics, even for $\mu<2$: a perfect agreement with 
the prediction
$\langle |X_t|^{\nu}\rangle \sim (\ln t)^{\nu/\mu}$ for $\nu<\mu$ is observed
in Fig. \ref{fig2}a. Importantly, the scaling function
$f(x)$ in this non-Markovian process is the same as for the underlying 
Markovian process between relocations (or with $q=0$). This property 
stems from the fact that the cumulant characteristic function 
$\ln \tilde{p}(k)$ [Eq. (\ref{exp})] and the function
$a(k)$ [Eq.(\ref{a})] have the same leading behavior at small $k$,
except for a multiplicative constant. In other words, 
the analyticity or non-analyticity of $\widetilde{P}(k,t)$
is {\it preserved} when $q$ is set different from zero. 

\section{Generalizations} 

We now show that several extensions of the PVM also admit a propagator
of the form given by Eq. (\ref{powerlaw}). 

\subsection{Decaying memory}

The results of the previous Section do not change qualitatively by 
considering memory kernels other than a pure preferential one. For instance,
the time in the past $t'$ may be chosen not uniformly like in Eq. (\ref{pref})
but with a probability decaying with $t-t'$,
the interval of time between 
a remembered occupation and the present time.
Consider, for instance, a power-law memory decay:
\begin{equation}\label{pigen}
\pi_t(t')=\frac{(t-t'+1)^{-\beta}}{\sum_{t''=0}^t(t-t''+1)^{-\beta}}
\end{equation}
with $\beta>0$ an exponent. Here, the visits are still
preferential, but with a tendency towards more recent sites 
(an effect actually observed in human mobility \cite{recency}).
If $\beta<1$ the sum in (\ref{pigen}) diverges at large $t$ and can be 
substituted by an integral; 
by taking the Fourier transform of (\ref{master}) and making the ansatz 
$\widetilde{P}(k,t)\simeq t^{-a(k)}$, one obtains an integral
equation for $a(k)$:
\begin{equation}\label{inta}
1-(1-q)\tilde{p}(k)=q(1-\beta)\int_0^1du(1-u)^{-\beta}u^{-a(k)}.
\end{equation}
Combining Eqs. (\ref{inta}) and (\ref{pk}) gives, at small $k$:
\begin{eqnarray}\label{agen}
&&a(k)\simeq\frac{1-q}{q}\ {\cal F}(\beta)C|k|^{\mu},\\
&&{\rm with}\quad {\cal F}(\beta)=\left[(1-\beta)
\int_0^1du(1-u)^{-\beta}\ln(1/u) \right]^{-1}.\nonumber
\end{eqnarray}
Eq. (\ref{agen}) shows that the scaling law (\ref{loglevy}) applies
to more general processes than the PVM.  
[Eq.(\ref{aasym}) is recovered for $\beta=0$.]
Interestingly, ${\cal F}(1)=\infty$, which indicates that the
scaling form (\ref{powerlaw}) breaks down 
for $\beta\ge1$. Actually, a similar calculation to the one above
shows that, for $\beta>2$, memory decays too fast to 
be relevant and the usual CLT (\ref{tcl}) is recovered. Of course,
these results do not mean that the aforementioned preservation property 
holds for arbitrary $\pi_t(t')$. For instance, for memory walks 
with $1<\beta<2$ and steps $\ell_i$ of finite variance,
the process is non-Gaussian \cite{romo}. Likewise, Brownian random
walks and L\'evy flights subjected to stochastic reseting
to the origin have asymptotic probability densities which are 
non-Gaussian \cite{evansmaj} and non-L\'evy \cite{nowak}, respectively.

\subsection{Model with bias} 
We now study the response of 
the non-Markovian walks (at fixed $q$) to the presence of a constant
forcing, namely, a bias
$\alpha\equiv\langle \ell\rangle=\sum_{\ell=-\infty}^{\infty}\ell p(\ell)
\ne 0$. Here,
we assume $\langle\ell^2\rangle<\infty$ or $\mu=2$.
By taking the first moment of Eq. (\ref{master}), an equation
for the average position 
$\langle X_t\rangle\equiv\sum_{n=-\infty}^{\infty}nP(n,t)$ is obtained:
\begin{equation}\label{eqmoment}
\langle X_{t+1}\rangle=(1-q)[\langle X_t\rangle+\alpha]
+q\sum_{t=0}^{t}\pi_t(t')\langle X_{t'}\rangle,
\end{equation}
for any kernel $\pi_t(t')$. We now denote
$\langle X_t^2\rangle_{_0}$ as the mean square displacement of the walker 
{\it at zero bias}. It is easy to show that 
$\langle X_t^2\rangle_{_0}$ obeys exactly the same equation as 
(\ref{eqmoment}), where $\alpha$ has to be replaced
by $\langle \ell^2\rangle_{_0}=
\sum_{\ell=-\infty}^{\infty}\ell^2 p_{_0}(\ell)$,
with $p_{_0}(\ell)$ unbiased. We deduce
an Einstein fluctuation-dissipation relation (FDR):
\begin{equation}\label{fdt}
\langle X_t\rangle=\frac{\alpha}{\langle \ell^2\rangle_{_0}}
\langle X_t^2\rangle_{_0}
\end{equation}
The exact equality (\ref{fdt}) is general: it is valid at all $t$ and 
for any kernel $\pi_t(t')$ (allowing to recover results
on the resetting to the origin with bias \cite{montero}). 
Despite of being out-of-equilibrium, the FDR with constant bias in this
system is the same as for 
ordinary random walks, where the response $\langle X_t\rangle$ is entirely 
determined by the fluctuations at zero bias.
With the kernel (\ref{pigen}) and $\beta<1$, the drift
is thus logarithmic: $\langle X_t\rangle\simeq
\alpha\frac{1-q}{q} {\cal F}(\beta)\ln t$, from 
Eqs. (\ref{fdt}) and (\ref{agen}) with $\mu=2$. The time evolution of
the first moment $\langle X_t\rangle$
is displayed in Fig.\ref{fig2}b-left for different parameter values. 

In other words, the effective friction coefficient
of the walker ($\propto\alpha\langle \dot{X}_t\rangle^{-1}$)
grows linearly with $t$. This illustrates the non-stationarity emerging from 
long range memory and the increasingly sluggish dynamics caused 
by frequent relocations to the same preferred sites.  

We further show that the combination of memory and 
bias has a drastic impact 
on the fluctuations of $X_t$ around $\langle X_t\rangle$. We 
take, for example, the PVM with n.n. steps 
in rule {\it (i)},
and expand Eq. (\ref{a}), which is valid for 
any $p(\ell)$, near $k=0$. Now using $\tilde{p}(k)=1-i\alpha k-\frac{1}{2}k^2+...$ 
we obtain $\widetilde{P}(k,t)\simeq
\exp[-i\mu_tk-\frac{1}{2}\sigma_tk^2]$, which corresponds for $P(n,t)$ to 
a Gaussian of mean $\mu_t$ and variance $\sigma_t$. We recover
$\mu_t=\alpha\frac{1-q}{q}\ln t$, see (\ref{fdt}), and obtain for $\sigma_t$:
\begin{equation}\label{sigma}
\sigma_t=\left[\frac{1-q}{q}+2\left(\frac{1-q}{q}\right)^2\alpha^2\right]
\ln t.
\end{equation}
If $q$ is small, the presence of a
bias therefore strongly {\it amplifies} the fluctuations of $X_t$, 
as the 2$^{\rm nd}$ term in (\ref{sigma}) is $>0$ and dominant. 
This effect is displayed in Fig. \ref{fig2}b-right. For ordinary n.n. 
random walks, on the contrary, the bias {\it decreases} the fluctuations: 
in that case $\sigma_t=(1-\alpha^2)t$ and motion 
becomes deterministic at $\alpha=1$ (see {\it e.g.} \cite{gleb}).

\section{Discussion and conclusion}

In summary, we have shown that L\'evy and Gaussian 
distributions can emerge generically far from the domain of applicability 
of the CLT, namely, in strongly subdiffusive path-dependent processes.
We emphasize that
the processes studied here exhibit subdiffusion because the relocation 
sites are selected 
heterogeneously in space. This situation is also encountered in the 
reseting to the origin, an extreme case where only one site 
receives all relocations, causing the typical diffusion 
length $l(t)$ to tend to a constant \cite{evansmaj}. 
To illustrate 
the importance of uneven relocations, one may by contrast consider
a n.n. random walk, which, in
rule {\it (ii)} above, relocates to a site chosen randomly 
and uniformly among the visited sites. In this case, $l(t)$ roughly obeys 
$dl/dt\sim (2R/l)[(R/2)/(1/q)]$, with $R=\sqrt{2D/q}$ the characteristic 
diffusion scale between two relocations, $2R/l$ being the probability of 
reseting near the edges of the territory covered by the walk. This leads to 
$l(t)\sim\sqrt{4Dt}$, a normal diffusive behavior, which is qualitatively
confirmed by the numerical simulations of Figure \ref{fig2}c.

The emergence of logarithmic diffusion can be understood qualitatively 
by drawing, from Fig. \ref{fig1}a,
an analogy with a branching random walk (see, {\it e.g.} 
\cite{biggins,branchsatya}). 
Consider an initial normal
random walk with a constant branching rate $q_b$. At each branching event,
a new random walk is created which starts from the current position of
the parent walk. The walks are independent, do not disappear, and all
branch at the same rate $q_b$. The process follows until it is stopped at some 
final time $T$. Let then imagine a single walker starting at the origin and 
following the paths left by all the branches, from the oldest to most recent, 
relocating at the start of the next branch when reaching the end of a branch.
The average number of branches at time $T$ is $N_{b}(T)=e^{q_b T}$ 
and the total number of steps needed for the single walker to walk 
along all of them is $t\simeq\int_0^{T}d\tau N_{b}(\tau)\simeq e^{q_bT}/q_{b}$. 
At time $t$, the single walker will be at a typical distance $l(t)$ from 
the origin, with  $l(t)^2\sim T\simeq \frac{1}{q_b}\ln t$. This form
is surprisingly similar to our result
$\langle X_t^2\rangle \simeq\frac{1}{q}\ln t$ for the PVM at small $q$. 
The argument above can be repeated 
with branching L\'evy flights, where $l(t)\propto T^{1/\mu}$, leading to 
a similar correspondence between the two models.

Note that the above analogy is only qualitative, as the PVM differs 
quantitatively from a set of branching RWs. 
Setting $q_b=q$, numerical simulations 
(not shown) indicate that, due to the rule of preferential visits, the 
relocation points in the memory model are distributed much more 
heterogeneously in space (namely, closer to the origin) than 
the branching points of the branching walks.

We conclude by mentioning that the processes studied here can explain two 
properties very often
observed in human and animal mobility \cite{gab,rhee,gonzalez,song,solis}: 
{\it a)} power-law distributed step lengths can coexist with a very slow 
diffusion in the 
long term ({\it i.e.}, home range behavior); {\it b)} the 
occupation of space by an individual within its home range is very non-uniform.
L\'evy flights with relocations to visited places are likely to be 
an efficient 
strategy for searching and exploiting renewable resources, a 
challenge faced by many living organisms 
\cite{gandhi,randomsearch,benichou,hills}.

\begin{acknowledgments}
We thank M. Marsili, O. Miramontes, I. Perez, 
J. R. Gomez-Solano and F. Sevilla for 
discussions. This work was supported by PAPIIT Grant IN105015,
by {\it Programa de Becas Posdoctorales en la UNAM}, and by the 
MPIPKS Advanced Study Group on 
Statistical Physics and Anomalous Dynamics of Foraging.
\end{acknowledgments}


\begin{thebibliography}{40}


\bibitem{feller} W. Feller, {\it An Introduction to Probability Theory
and its Applications}, Vol. 2 (Wiley, New York, 2008).

\bibitem{alpha} G. Samoradnitsky and M. S. Taqqu,
{\it Stable Non-Gaussian Random Processes: Stochastic Models with Infinite 
Variance} (Chapman $\&$ Hall, 1994).

\bibitem{bardou} F. Bardou, J.-P. Bouchaud, A. Aspect, and C. Cohen-Tannoudji,
{\it L\'evy Statistics and Laser Cooling} (Cambridge, 2002).

\bibitem{light} P. Barthelemy, J. Bertolotti, and  D. S. Wiersma,
Nature {\bf 453}, 495 (2008).

\bibitem{strangekin} M. F. Shlesinger, G. M. Zaslavsky, and J. Klafter,
Nature {\bf 363}, 31 (1993).

\bibitem{bouchaud} J.-P. Bouchaud and A. Georges, 
Phys. Rep. {\bf 195}, 127 (1990).

\bibitem{rainer} A. V. Chechkin, R. Metzler, J. Klafter, and V. Yu. Gonchar,
in {\it Anomalous Transport: Foundations and Applications},
edited by R. Klages, G. Radons, and I. M. Sokolov (Wiley-VCH Verlag,
Weinheim, 2008), pp. 129--159.

\bibitem{klafterphysrep} R. Metzler and J. Klafter, 
Phys. Rep. {\bf 339}, 1 (2000).

\bibitem{weiss} G. H. Weiss, {\it Aspect and Applications of the Random
Walk} (Elsevier, Amsterdam, 1994).

\bibitem{lw} V. Zaburdaev, S. Denisov, and J. Klafter,
Rev. Mod. Phys. {\bf 87}, 483 (2015).

\bibitem{tcells} T. H. Harris {\it et al.}, Nature {\bf 486}, 545 (2012).

\bibitem{gandhi} G. M. Viswanathan, S. V. Buldyrev, S. Havlin, M. G. E. 
da Luz, E. P. Raposo, and H. E. Stanley, {\it Nature} {\bf 401}, 911 (1999).

\bibitem{levy2}
F. Bartumeus, M. G. E. da Luz, G. M. Viswanathan, and J. Catalan,
Ecology {\bf 86}, 3078 (2005).

\bibitem{sims}
D. W. Sims {\it et al.}, Nature {\bf 451}, 1098 (2008).

\bibitem{gab}
G. Ramos-Fern\'andez, J. L. Mateos, O. Miramontes, G. Cocho, H. Larralde, and
B. Ayala-Orozco, Behav. Ecol. Sociobiol. {\bf 55}, 223 (2004).

\bibitem{brown} C. T. Brown, L. S. Liebovitch, and R. Glendon,
Hum. Ecol. {\bf 35}, 129 (2007).

\bibitem{pnas2014} D. A. Raichlen, B. M. Wood, A. D. Gordon, A. Z. P. Mabulla, 
F. W. Marlowe, and H. Pontzer, Proc. Natl. Acad. Sci. USA {\bf 111}, 728 
(2014).

\bibitem{geisel}
D. Brockmann, L. Hufnagel, and T. Geisel,
Nature {\bf 439}, 462 (2006).

\bibitem{rhee} I. Rhee, M. Shin, S. Hong, K. Lee, and  S. Chong,
In Proc. IEEE INFOCOM, 13-18 April 2008, Phoenix, AZ
(IEEE, Piscataway, NJ, 2008), pp. 924-932.

\bibitem{gonzalez}
M. C. Gonz\'alez, C. A. Hidalgo, and A.-L. Barab\'asi, 
Nature {\bf 453}, 779 (2008).

\bibitem{song} 
C. Song, T. Koren, P. Wang, and A.-L. Barab\'asi, 
Nature Phys. {\bf 6}, 818 (2010).

\bibitem{gautestad2005}
A. O. Gautestad and I. Mysterud, Am. Nat. {\bf 165}, 44 (2005).

\bibitem{gautestad2006}
A. O. Gautestad and I. Mysterud, Ecol. Complex. {\bf 3}, 44 (2006).


\bibitem{borger}
L. B${\rm\ddot{o}}$rger, B. D. Dalziel, and J. M. Fryxell, 
Ecol. Lett. {\bf 11}, 637 (2008). 

\bibitem{solis} D. Boyer and C. Solis-Salas, Phys. Rev. Lett.
{\bf 112}, 240601 (2014).


\bibitem{kantz} 
S. I. Denisov and H. Kantz,
Phys. Rev. E {\bf 83}, 041132 (2011).

\bibitem{drager}
J. Dr${\rm \ddot{a}}$ger and J. Klafter, Phys. Rev. Lett. {\bf 84},
5998 (2000); R. Venegeroles, J Stat Phys {\bf 154}, 988 (2014).

\bibitem{sokolov}
J. Klafter and I. M. Sokolov, {\it First Steps in Random Walks}
(Oxford, Oxford, 2011).

\bibitem{fagan}
W. F. Fagan {\it et al.}, Ecol. Lett. {\bf 16}, 1316 (2013).

\bibitem{vanmoorter}
B. van Moorter {\it et al.}, Oikos {\bf 118}, 641 (2009). 

\bibitem{compet} M. Magdziarz and A. Weron, 
Phys. Rev. E {\bf 75}, 056702 (2007).

\bibitem{hilhorst} H. J. Hilhorst, Braz. J. Phys. {\bf 39}, 371 (2009).

\bibitem{elephant}
G. M. Sch${\rm \ddot{u}}$tz and S. Trimper,
Phys. Rev. E {\bf 70}, 045101(R) (2004).

\bibitem{kumar} N. Kumar, U. Harbola, and K. Lindenberg,
Phys. Rev. E {\bf 82}, 021101 (2010).

\bibitem{dasilva} M. A. A. da Silva, G. M. Viswanathan, and J. C. Cressoni,
Phys. Rev. E {\bf 89}, 052110 (2014).

\bibitem{davis} B. Davis, Probab. Theor. Related Fields {\bf 84}, 203 (1990).

\bibitem{annals} E. Bolthausen and U. Schmock, Ann. Probab. {\bf 25}, 531 (1997).

\bibitem{quasistatic}
V. B. Sapozhdcov, J. Phys. A: Math. Gen. {\bf 27}, L151 (1994).

\bibitem{grassberger}
J. G. Foster, P. Grassberger, and M. Paczuski, 
New J. Phys. {\bf 11}, 023009 (2009).

\bibitem{siam}
H. G. Othmer and A. Stevens,
SIAM J. Appl. Math. {\bf 57}, 1044 (1997).

\bibitem{sbm} S. C. Lim and S. V. Muniandy, 
Phys. Rev. E {\bf 66}, 021114 (2002).

\bibitem{slowsbm} A. S. Bodrova, A. V. Chechkin, A. G. Cherstvy, and R. Metzler,
New J. Phys. {\bf 17}, 063038 (2015).

\bibitem{matteo} M. Vendruscolo and M. Marsili, 
Phys. Rev. E {\bf 54}, R1021 (1996).

\bibitem{schmiedeberg} M. Schmiedeberg, V. Y. Zaburdaev, and H. Stark,
J. Stat. Mech. P12020 (2009).

\bibitem{fleurov} E. Barkai, V. Fleurov, and J. Klafter, Phys. Rev. E 
{\bf 61}, 1164 (2000).
 
\bibitem{romo} D. Boyer and  J. C. R. Romo-Cruz,
Phys. Rev. E {\bf 90}, 042136 (2014).

\bibitem{evansmaj} M. R. Evans and S. N. Majumdar,
Phys. Rev. Lett. {\bf 106}, 160601 (2011);
J. Phys. A: Math. Theor. {\bf 44,} 435001 (2011).

\bibitem{montero} M. Montero and J. Villarroel,
Phys. Rev. E {\bf 87}, 012116 (2013).

\bibitem{resetlevy} L. Ku\'smierz, S. N. Majumdar, S. Sabhapandit, and 
G. Schehr, Phys. Rev. Lett. {\bf 113}, 220602 (2014).

\bibitem{nowak} L. Ku\'smierz and E. Gudowska-Nowak,
Phys. Rev. E {\bf 92}, 052127 (2015).

\bibitem{recency} H. Barbosa, F. Buarque de Lima Neto, A. Evsukoff, 
and R. Menezes,
arXiv:1504.01442 [physics.soc-ph] (2015).

\bibitem{gleb} O. B\'enichou, K. Lindenberg, and G. Oshanin,
Phys. A {\bf 392}, 3909 (2013). 

\bibitem{biggins}
J.D. Biggins, Stochastic Process. Appl. {\bf 34}, 255 (1990).

\bibitem{branchsatya}
K. Ramola, S. N. Majumdar, and G. Schehr,
Chaos Soliton. Fract. {\bf 74}, 79 (2015).

\bibitem{benichou} O. B\'enichou, C. Loverdo, M. Moreau, and R. Voituriez,
Rev. Mod. Phys. {\bf 83}, 81 (2011).

\bibitem{randomsearch}
G. M. Viswanathan, M. G. E. da Luz, E. P. Raposo, and H. E. Stanley,
{\it The Physics of foraging} (Cambridge, Cambridge, 2011).

\bibitem{hills} T. T. Hills, P. M. Todd, D. Lazer,
A. D. Redish, and I. D. Couzin, Trends Cogn. Sci. {\bf 19}, 46 (2015).




%
%




\end{thebibliography}
\end{document}